\documentclass[twocolumn,superscriptaddress,nofootinbib,amsmath,amssymb]{aastex63}

\usepackage{times}
\usepackage{amsmath}

\newcommand{\RN}[1]{\uppercase\expandafter{\romannumeral#1}}

\newcommand{\eg}{e.g.,~}
\newcommand{\ie}{i.e.,~}

\shorttitle{Crustal magnetic fields do not lead to large magnetic-field amplifications}
\shortauthors{Chabanov, Tootle, Most and Rezzolla}
\graphicspath{{./}{figures/}}

\begin{document}


\title{Crustal magnetic fields do not lead to large magnetic-field amplifications in
  binary neutron-star mergers}

\author[0000-0001-9676-765X]{Michail Chabanov}
\affiliation{Institut f\"ur Theoretische Physik, Goethe Universit\"at,
Max-von-Laue-Str. 1,
60438 Frankfurt am Main, Germany}

\author[0000-0001-9781-0496]{Samuel D. Tootle}
\affiliation{Institut f\"ur Theoretische Physik, Goethe Universit\"at,
Max-von-Laue-Str. 1,
60438 Frankfurt am Main, Germany}

\author[0000-0002-0491-1210]{Elias R. Most}
\affiliation{Princeton Center for Theoretical Science, 
Princeton University, 
Princeton, NJ 08544, USA}
\affiliation{Princeton Gravity Initiative,
Princeton University,
Princeton, NJ 08544, USA}
\affiliation{School of Natural Sciences, 
Institute for Advanced Study, 
Princeton, NJ 08540, USA}

\author[0000-0002-1330-7103]{Luciano Rezzolla}
\affiliation{Institut f\"ur Theoretische Physik, Goethe Universit\"at,
Max-von-Laue-Str. 1,
60438 Frankfurt am Main, Germany}
\affiliation{Frankfurt Institute for Advanced Studies,
Ruth-Moufang-Str. 1,
60438 Frankfurt am Main, Germany}
\affiliation{School of Mathematics, Trinity College,
Dublin 2, Ireland}



\begin{abstract}
The amplification of magnetic fields plays an important role in
explaining numerous astrophysical phenomena associated with binary
neutron-star mergers, such as mass ejection and the powering of short
gamma-ray bursts. Magnetic fields in isolated neutron stars are often
assumed to be confined to a small region near the stellar surface, while
they are normally taken to fill the whole stars in the numerical
modelling. By performing high-resolution, global, and high-order
general-relativistic magnetohydrodynamic simulations we investigate the
impact of a purely crustal magnetic field and contrast it with the
standard configuration consisting of a dipolar magnetic field with the
same magnetic energy but filling the whole star. While the
crust-configurations are very effective in generating strong magnetic
fields during the Kelvin-Helmholtz-instability stage, they fail to
achieve the same level of magnetic-field amplification of the full-star
configurations. This is due to the lack of magnetized material in the
neutron-star interiors to be used for further turbulent amplification and
to the surface losses of highly magnetized matter in the
crust-configurations. Hence, the final magnetic energies in the two
configurations differ by more than one order of magnitude. We briefly
discuss the impact of these results on astrophysical observables and how
they can be employed to deduce the magnetic topology in merging
binaries.
\end{abstract}

\keywords{stars: neutron --- stars: magnetic field ---
magnetohydrodynamics (MHD) --- methods: numerical}


\section{Introduction}
\label{sec:intro}

The coincident detection of the gravitational-wave signal GW170817, of
the short gamma-ray burst GRB 170817A and of the kilonova
AT2017gfo~\citep{Abbott2017_etal, Drout2017, Cowperthwaite2017} has
provided strong evidence that short gamma-ray bursts are related to
binary neutron-star (BNS) mergers and that magnetic fields play an
important role in the post-merger evolution. This conclusion stems from a
series of studies identifying magnetic fields as crucial to generate
magnetically driven outflows \citep{Kiuchi2012b, Siegel2013, Kiuchi2017,
  Fernandez2018,Ciolfi2020_a,Fujibayashi2023}, but also to produce the
conditions necessary for jet formation and launching ~\citep{Liu:2008xy,
  Anderson2008, Rezzolla:2011, Palenzuela2013a, Kiuchi2015,
  Murguia-Berthier2016,Ciolfi2020c, Nathanail2020b, Nathanail2020c,
  Gottlieb2022}.

Early studies identified that turbulence, starting from the
Kelvin-Helmholtz instability (KHI) triggered in the first few
milliseconds after merger by the shearing of the stellar surfaces, is
essential in the amplification of the magnetic field \citep{Price06,
  Giacomazzo:2010, Kiuchi2015a, Aguilera-Miret2020}. However, because the
accurate description of the turbulent motions can only be achieved at
enormous computational cost, a number of suggestions have been made over
the years to obtain a subgrid-scale modeling of the magnetic-field
evolution. These so-called ``large-eddy simulations'' (LESs) have
attempted to incorporate the small-scale, dynamo-driven, magnetic-field
amplification into \textit{global} high-resolution general-relativistic
magnetohydrodynamics (GRMHD) simulations employing computationally
affordable resolutions~\citep{Giacomazzo:2014a, Palenzuela2015,
  Radice2017, Aguilera-Miret2020}. At the same time, \textit{local} and
high-resolution simulations of KHI-driven turbulence have shown evidence
of possible converged and saturated magnetic-field amplification
\citep[see, \eg][]{Zrake2013b, Obergaulinger10}.

While a generally accepted view of the process of magnetic-field
amplification is still missing, broad consensus is present that the
maximum achievable magnetic-field strength is reached by an equipartition
between the magnetic energy and the kinetic energy on the smallest
scales, thus yielding ``magnetar-strength'' fields of the order
$\gtrsim10^{16}\,\mathrm{G}$~\citep[see, \eg][for some initial
  estimates]{Price06, Liu:2008xy, Giacomazzo:2009mp}. This expectation,
which has so far been difficult to demonstrate via direct global
simulations, has recently been shown to hold at least with
high-resolution LES simulations by \citet{Palenzuela_2022PRD}, where the
amplification of the averaged magnetic field saturated to approximately
$10^{16}\,\mathrm{G}$.

A distinct but equally interesting question is that regarding the role
played by the initial magnetic-field strength and
topology~\citep[see][for some initial investigations]{Giacomazzo:2009mp,
  Kawamura2016, Ruiz2020a}. Following a long list of works in this
area,~\citet{Aguilera-Miret2021} have recently found that the initial
magnetic-field topology is quickly destroyed and that the final turbulent
state depends only weakly on the configuration considered, \eg dipoles
with different strengths, misaligned dipoles, and even a multipolar
structure. A common thread in all of these studies is the assumption that
the magnetic fields are confined to the neutron stars and permeating the
whole stellar structure. Yet, a number of works exploring the rotational,
thermal, and magnetic evolution of neutron stars have considered as most
natural those configurations in which the magnetic field is concentrated
only in the crust \citep[see, \eg][]{Pons2007, Pons2009}. This
configuration follows from assuming the stellar core as a type-\RN{1}
superconductor that expels most of its magnetic flux on a very short
timescale~\citep[see also][]{Vigano2021}. Here, we explore a BNS scenario
in which, by the time of the merger, the magnetic fields in the stellar
cores have been expelled or have decayed, so that only crustal fields are
present. If such configurations are indeed common in neutron stars, and
since the stellar crusts play a fundamental role in the development of
the KHI, it is interesting to assess whether strong crustal magnetic
fields lead to a distinct amplification process and to a different
gravitational-wave signal.

To address this question, we report the results of high-resolution,
global, and high-order GRMHD simulations of merging BNSs with two
different initial magnetic-fields topologies. More specifically, in
addition to the standard configuration with a dipolar magnetic field
filling the whole star, we consider a configuration where the magnetic
field is still dipolar and has the \textit{same} total magnetic energy,
but is confined to the crust only. Overall, our simulations indicate that
such ``crust-configurations'' fail to produce a sufficiently large
turbulent amplification of the magnetic field and lead to a post-merger
hypermassive neutron star (HMNS) with an amplified electromagnetic energy
that is $\lesssim 5 \%$ of the corresponding ``full-configurations''.

\section{Mathematical and Numerical Methods}
\label{sec:methods}

The simulations reported below are obtained after solving the Einstein
equations together with those of ideal GRMHD via the high-order
high-resolution shock-capturing code \texttt{FIL}~\citep{Most2019b,
  Most2019c} and the temperature-dependent equation of state
\texttt{TNTYST}, which has a maximum mass of $M_{\mathrm{TOV}} =
2.21~M_{\odot}$~\citep{Togashi2017}. To assess whether the resolution
employed is sufficient to capture the relevant physical behaviour, we
employ two different resolutions on the highest (7th) refinement level
with $\Delta x\sim 0.047~M_{\odot}\approx 70\,\mathrm{m}$, or of $\Delta
x\sim 0.071~M_{\odot}\approx 105\,\mathrm{m}$; we note that although
these are not the highest resolutions employed so far in the
literature~\citep[see, \eg][]{Siegel2013, Kiuchi2017}, they benefit from
the use of a high-order code~\citep{Most2019b}. Hereafter, we will refer
to these two setups respectively as high (HR) and low (LR) resolutions,
but discuss the results of the former only; details on the LR results and
their differences with the HR ones are presented in the Appendix. We note
that although the resolutions employed here are very high, the
magnetorotational instability (MRI) is normally under-resolved for most
of the matter in our simulations as estimated via different MRI quality
factors~\citep{Siegel2013, Kiuchi2017}. However, we expect the MRI
  to have a little impact on the results obtained here and for two
  different reasons. First, the MRI would develop only in a region where
  $\partial_{\varpi} \Omega < 0$, where $\Omega$ and $\varpi$ are
  respectively the angular velocity and the distance from the rotation
  axis; however the large shear produced at the merger and the subsequent
  turbulence that follows prevent the development of a coherent angular
  velocity profile in the merged object at least initially. Second,
  because even when a coherent angular velocity profile is developed, the
  MRI would have the same impact on both configurations since the region
  where $\partial_{\varpi} \Omega < 0$ is very similar and with
  comparable magnetic-field strengths in the two cases.
 
\begin{figure*}
\includegraphics[width=0.475\textwidth,height=0.40\textwidth]{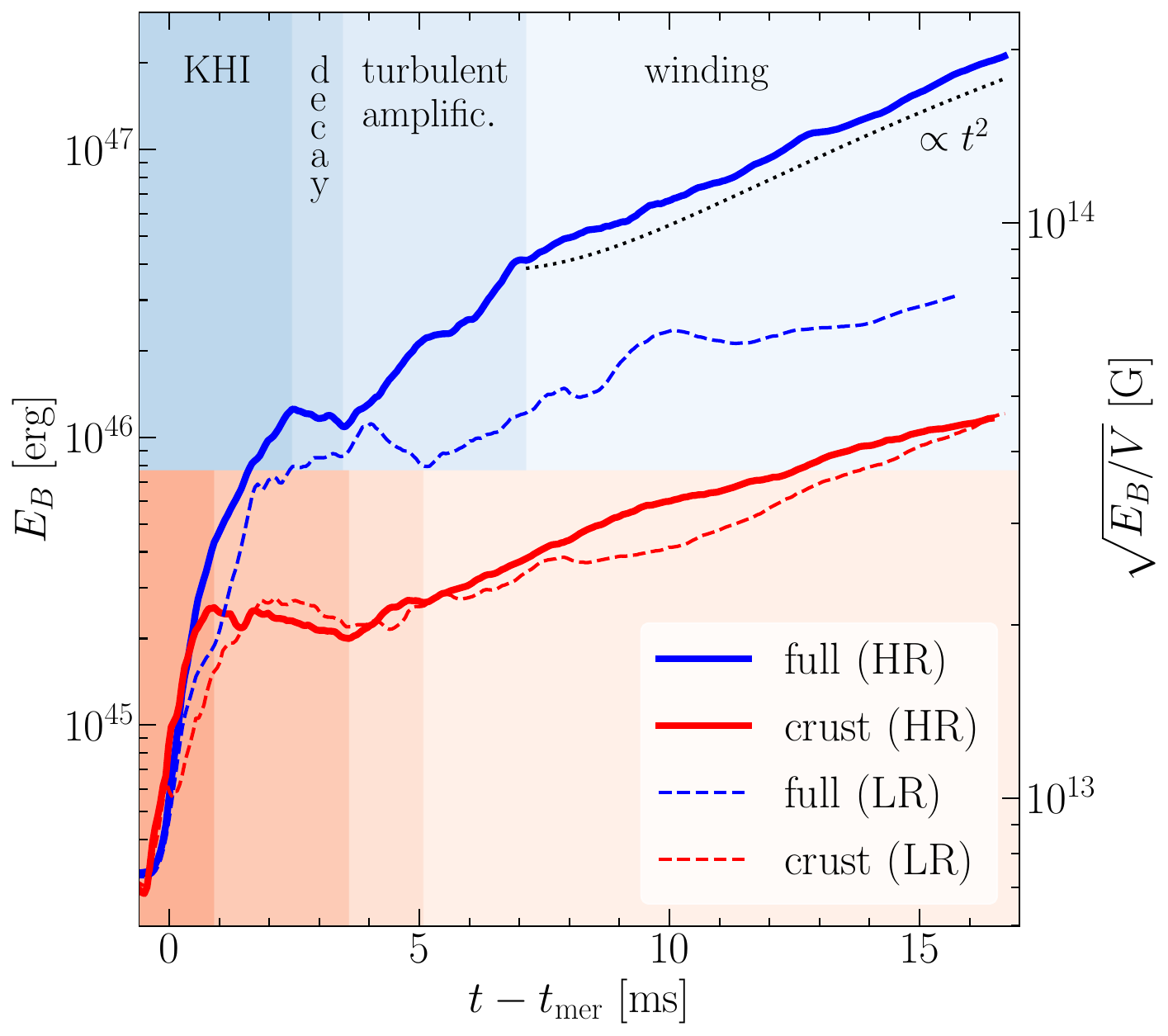}
\hskip 0.5cm
\includegraphics[width=0.475\textwidth,height=0.40\textwidth]{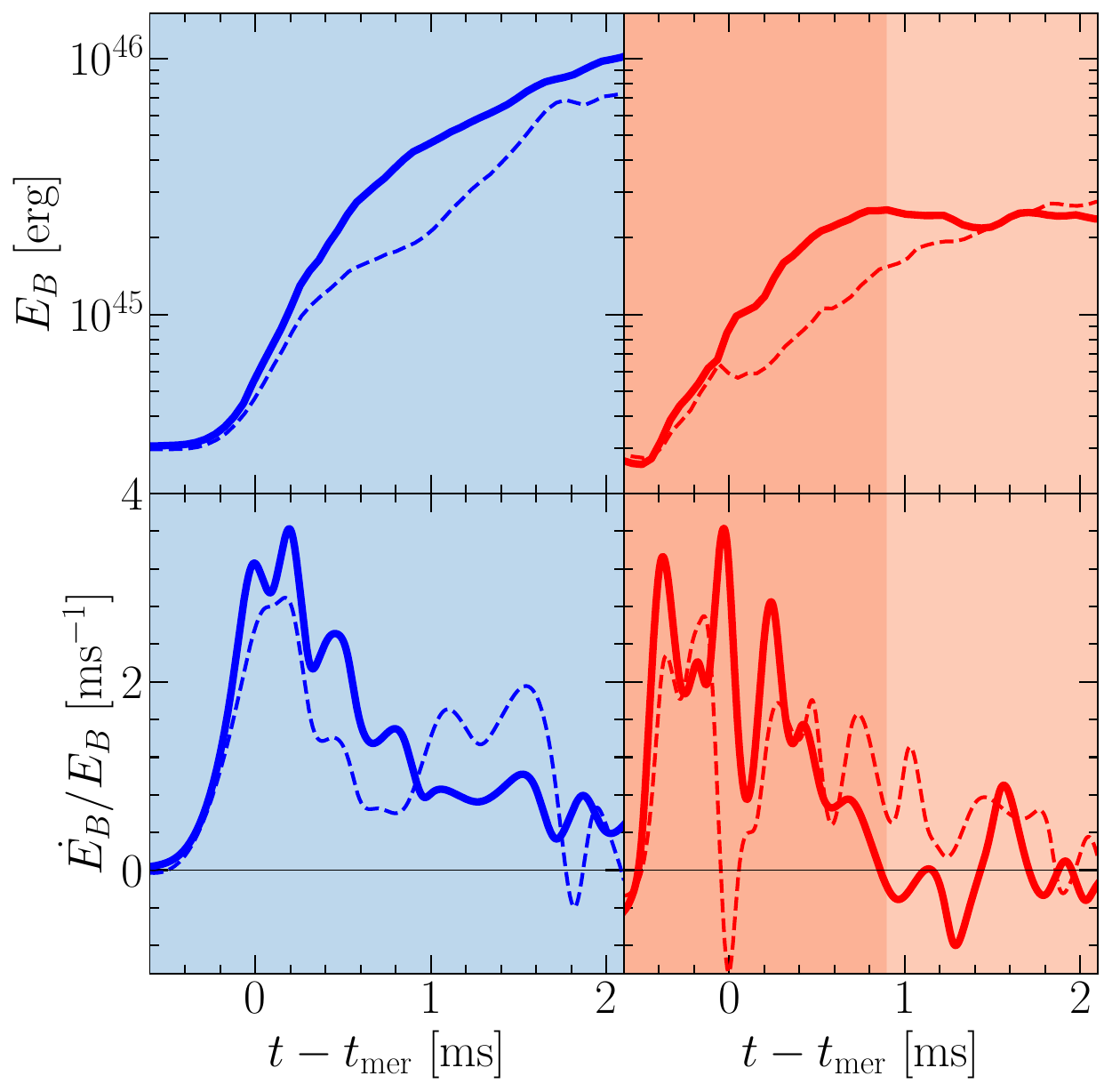}
\caption{\textit{Left:} Evolution of the total magnetic energy $E_B$ for
  the high-resolution (HR) full- (blue solid line) and
  crust-configuration (red solid line); dashed blue (red) lines refer to
  the low-resolution (LR) full- and crust-configuration,
  respectively. Different shadings highlight the four stages of the
  evolution. The black dotted line shows a quadratic fit to full
    (HR) in the winding stage and is shifted downwards for better
    visualization. The quantity $V$ denotes the reference volume of a
  sphere with a radius of $11\,\mathrm{km}$. \textit{Right top}: The same
  as in the left but for $t-t_{\mathrm{mer}}\lesssim2\,\mathrm{ms}$.
  \textit{Right bottom}: the same as the top but for the growth-rate
  $\dot{E}_B/E_B$.}
\label{fig:total_energy}
\end{figure*}

In both the ``full-'' and ``crust-configurations'', the magnetic-field is
initialized via the vector potential as $A_{i} =
A_b\,[-(x^j-x^j_{_{\mathrm{NS}}}) {\epsilon}_{ij}] \mathrm{exp}
[-g_w(r-g_r)^2] \mathrm{max}(p-p_{{\rm co}},0)^{n}$ for $i=x,y$ and
$A_z=0$, where $x^j_{_\mathrm{NS}}$ denote the coordinate centers of the
two stars ($x^3_{_{\mathrm{NS}}}=z_{_{\mathrm{NS}}}=0)$, $r:=
\sqrt{\delta_{ij}(x^i - x^i_{_{\mathrm{NS}}}) (x^j -
x^j_{_{\mathrm{NS}}})}$, and ${\epsilon}_{ij}$ is the Levi-Civita symbol.
The magnetized region extends outwards to the radius where the pressure
reaches the cut-off pressure $p_{{\rm co}} \approx 6\times10^{-5}\times
p_{{c}}$, where $p_{{c}}$ is the central pressure, and inwards to the
radius $R_{\rm in}$, such that $A_{i} = 0$ if $r<R_{\rm in}$ (see Table
\ref{tab:models} in the Appendix for additional details on the seeding
parameters).

With such a prescription, the initial magnetic field has closed loops
around a neutral line that is at $\simeq 0.52~R_{\rm NS}$ for the
full-configurations, while it is at $\simeq 0.87~R_{\rm NS}$ for the
crust-configurations (a view of the magnetic-field topology soon
  before merger is shown in Fig. \ref{fig:initial} in the Appendix). For
the high (low) resolutions, the magnetized crust is resolved with
$\gtrsim 17$ ($\gtrsim 11$) gridpoints and does not suffer from
significant numerical dissipation in the remainder of the
inspiral. Finally, we note that to reduce diffusion during the inspiral,
the magnetic fields are seeded when the separation between the stars
decreases below $8.9~M_{\odot}$, which corresponds to
$t-t_{\mathrm{mer}}\lesssim-1\,\mathrm{ms}$.

\section{Results}
\label{sec:results}

We start by describing the overall dynamics with particular focus on the
difference between the full and crust-configurations relative to the
magnetic-field amplification. In the left panel of
Fig.~\ref{fig:total_energy} we report the evolution of the total
electromagnetic energy in the whole computational domain
  $\mathcal{V}$, $E_B:=\int_\mathcal{V} \mathrm{d}^3x \sqrt{\gamma}
(E^2+B^2)/8\pi$, where $\gamma$ is the determinant of the three-metric,
and $B^2:=B_i B^i~(E^2:=E_i E^i)$ is the square modulus of the magnetic
(electric) field in the Eulerian frame.  The different lines refer to the
crust- (red) and full-configurations (blue), with dashed and solid lines
reporting the behaviour of the low- (LR) and high-resolution (HR)
simulations, respectively. Note that all configurations start with the
\textit{same} electromagnetic energy of $\sim 3 \times
10^{44}\,\mathrm{erg}$, corresponding to a maximum initial magnetic-field
strength of $\sim 10^{14}\,\mathrm{G}$ ($\sim 2.4 \times
10^{14}\,\mathrm{G}$) for the full-configuration (crust-configuration;
see also Table \ref{tab:models}).

\begin{figure*}
\includegraphics[width=0.99\textwidth]{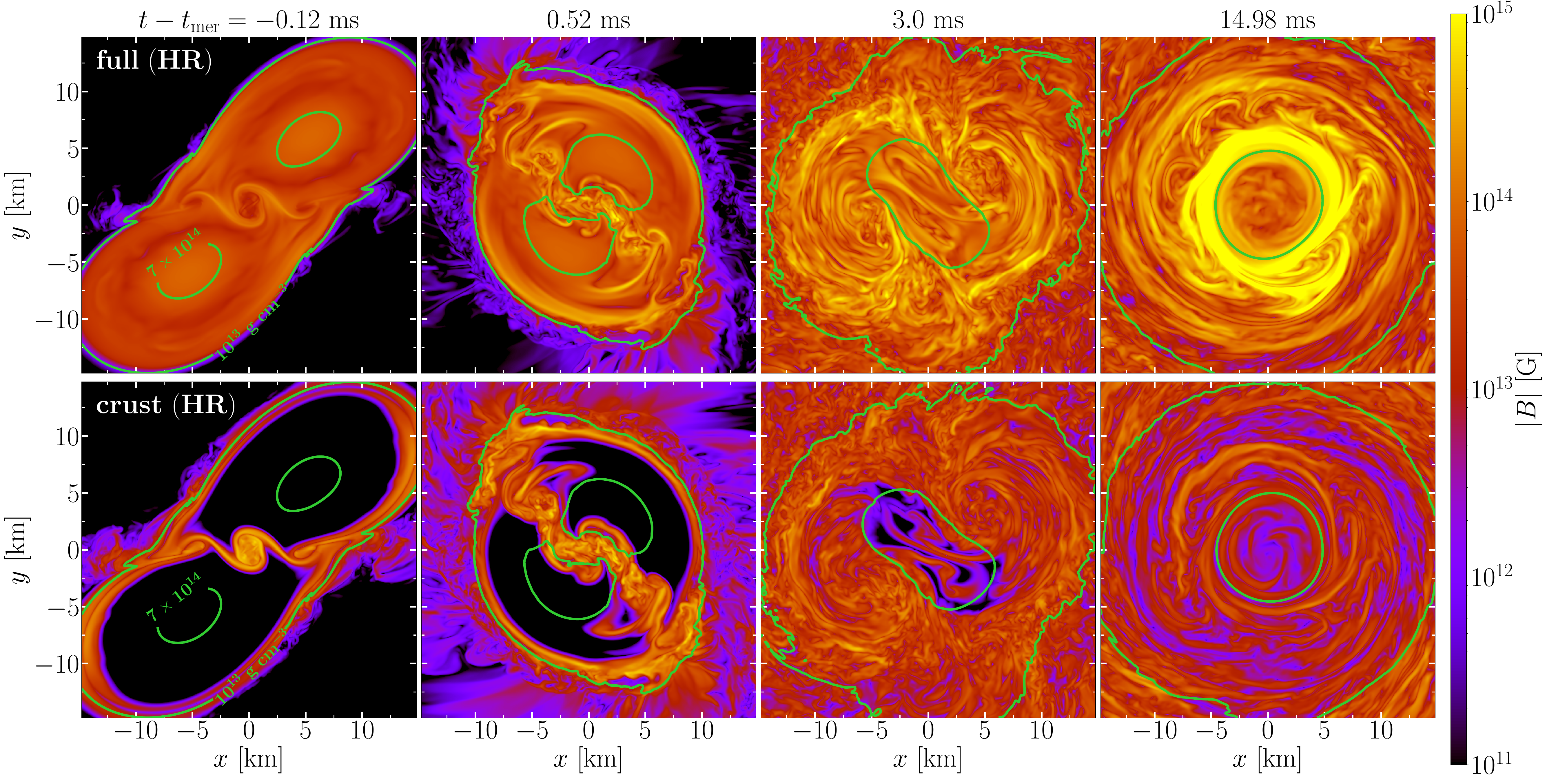}
\caption{Distributions on the $(x,y)$ plane of the magnetic-field
  strength $|B|$ from the HR simulations at four representative
  times. The top and bottom rows refer to the full- and
  crust-configurations, respectively.}
\label{fig:normB}
\end{figure*}

To describe such a complex dynamics, it is convenient to classify the
evolution of $E_B$ in four distinct stages that are highlighted by
different shadings in Fig.~\ref{fig:total_energy}. The first stage of the
evolution, or ``KHI-driven stage'', begins when tidal forces start
significantly deforming both stars, at $t-t_{\mathrm{mer}}\approx
-0.6\,\mathrm{ms}$ ($-0.6$), and ends when the KHI-driven turbulence
ceases to increase $E_B$ at $t-t_{\mathrm{mer}} \approx
2.46\,\mathrm{ms}$ ($0.9$) for the full- (crust-)
configuration\footnote{A more careful examination of
Fig.~\ref{fig:total_energy} and \ref{fig:normB} reveals that there are
two different phases in the KHI-driven stage with slightly different
growth-rates; for simplicity, we will ignore these small
differences.}. The KHI-unstable shear layer can be seen in the first
column of Fig.~\ref{fig:normB}, which presents cross-sections of the norm
of the magnetic field $|B|:=\sqrt{B^{i}B_{i}}$ in the $(x,y)$-plane at
time $t-t_{\mathrm{mer}}=-0.12\,\mathrm{ms}$ and at an
elevation\footnote{We use this elevation to minimise the influence of the
boundary conditions across the equatorial plane; smaller elevations yield
very similar views.}  of $z \simeq 1 ~ \mathrm{km} \lesssim 0.115~R_{\rm
  NS}$ for the full- (top row) and crust-configuration (bottom row). The
second column of Fig.~\ref{fig:normB}, at
$t-t_{\mathrm{mer}}=0.52\,\mathrm{ms}$, highlights how the KHI-driven
turbulent motion at the interface between the stars extends to other
regions and involves also the stellar interiors. At this point, the
amplification of the magnetic field has almost ended for the
crust-configuration, as can be appreciated from the left panel of
Fig.~\ref{fig:total_energy}.

The second stage of the evolution, or ``decay stage'', is instead
characterized by a lack of amplification and the evolution of $E_B$ is
constant in time or exhibits a weak decay. This is the result of a
combination of factors: the expansion of the dense HMNS core ($\rho
  \gtrsim 10^{13}\,{\rm gm/cm}^3$), the ejection of matter at the HMNS
surface, and an insufficient magnetic-field amplification. During
  this stage, magnetic pressure gradients accelerate the expansion of
  matter in the remnant and hence are doing work on the fluid. At the
  same time, the KHI-driven turbulent motion ceases to be produced
  efficiently as the kinetic energy in the remant is reduced through the
  emission of gravitational waves, shock-heating and matter
  ejection. This leads to an overall decrease in the electromagnetic
  energy. It is interesting to note that during this stage, the lack of
turbulence in the HMNS core results in the magnetic fields of the
crust-configurations to remain very weak and the full-configuration to
preserve its initial large-scale coherence; this can be seen in the third
column of Fig.~\ref{fig:normB} at $t-t_{\mathrm{mer}} = 3\,\mathrm{ms}$.
The decay stage ends for the full- (crust-) configuration at
$t-t_{\mathrm{mer}}=3.48\,\mathrm{ms}$ ($3.59$).

The third stage, or ``turbulent-amplification stage'', is characterized
by a new process of magnetic-field amplification that becomes strong
enough to counteract the decay of the previous stage and leads to a net
amplification~\citep[see also the high-resolution calculations
  of][]{Siegel2013}. At this point in the evolution, the HMNS has emerged
from the highly nonlinear early post-merger phase in which the two
stellar cores collide and bounce~\citep[see][for a toy
  model]{Takami2015}. Subsequent turbulent motions can develop and lead
to a nonlinear amplification of the magnetic field -- with a smaller
growth-rate than in the KHI stage and not related to the development of
the MRI -- that result into a substantial growth of the poloidal field up
to $t-t_{\mathrm{mer}} = 7.13\,\mathrm{ms}$ ($5.09$).

The fourth and final stage, or ``winding stage'', starts when the
turbulence is fully developed and more regular, large-scale shearing
motions can be produced in the HMNS. Under the infinite-conductivity
conditions of ideal-MHD, these motions lead to the well-known
linear-in-time growth of the magnetic field that represents the
quasi-stationary state reached in our simulations (see the final part of
the Fig.~\ref{fig:total_energy} and the fourth column of
Fig.~\ref{fig:normB}). Obviously, the winding stage cannot continue
indefinitely, but will terminate when the magnetic-field energy is in
equipartition with the kinetic energy stored in the differential
rotation, so that further amplification is energetically disfavoured. We
should note that the classification discussed here describes well the
dynamics in our simulations, which have moderate initial magnetic fields.
However, if the initial exponential amplification is much larger -- as a
result of additional driving terms in the MHD
equations~\citep{Giacomazzo:2014a, Palenzuela_2022PRD} or of very large
initial magnetic fields~\citep{Kiuchi2017} -- equipartition may be
reached much earlier and the subsequent stages may be
absent~\citep[see][where these stages are not found]{Kiuchi2017,
  Giacomazzo:2014a, Palenzuela_2022PRD}.

\begin{figure*}
\centering
\includegraphics[width=0.99\textwidth]{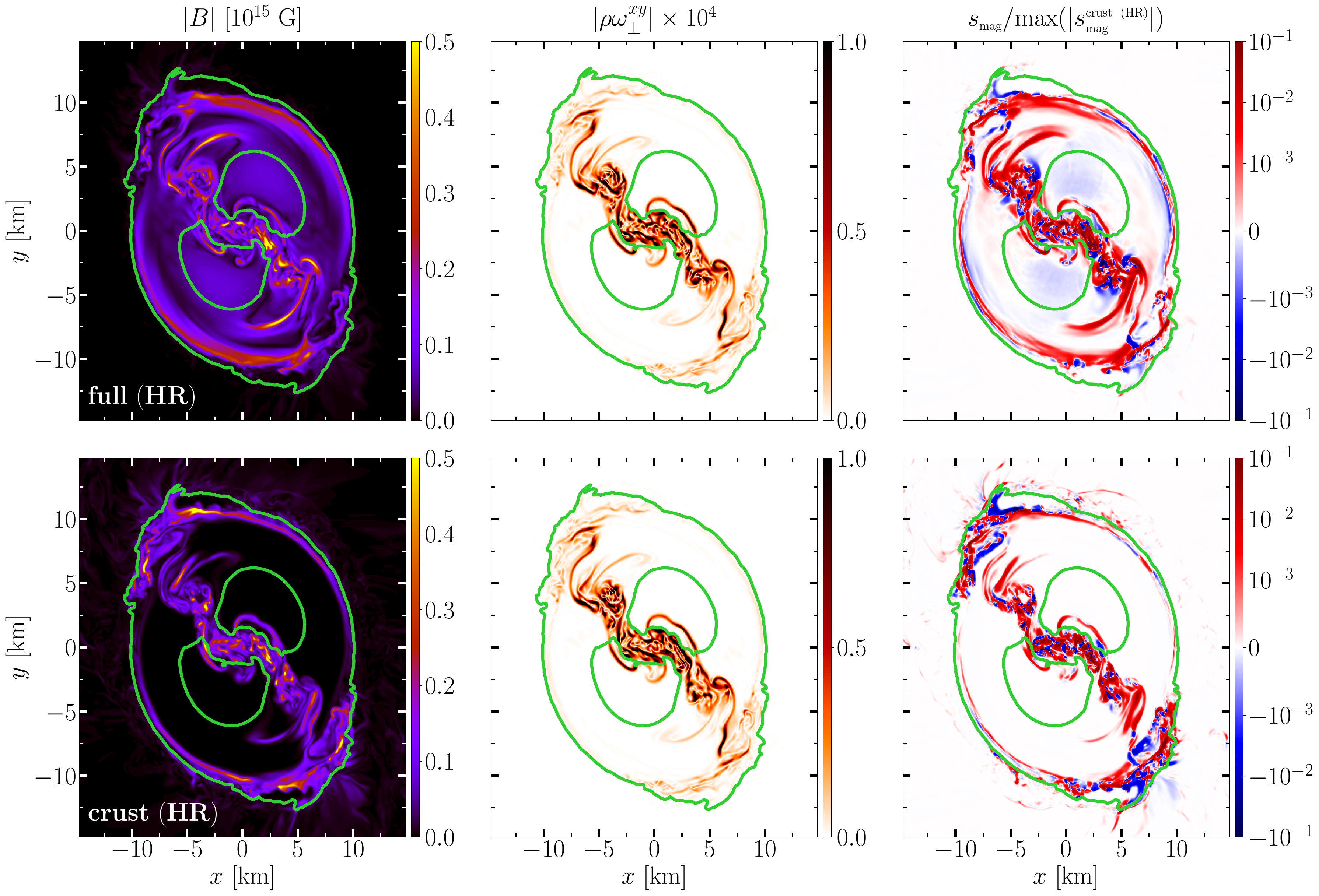}
\caption{Distributions on the $(x,y)$ plane of the magnetic-field
  strength $|B|$ (left column), of the density-weighted $xy$-component of
  the kinematic vorticity $|\rho \omega^{xy}_{\perp}|$ (middle column),
  and of the amplification source term $s_{\mathrm{mag}}$ (right column)
  at $t-t_{\mathrm{mer}}=0.52~\mathrm{ms}$. The top and bottom rows refer
  to the full- and crust-configurations, respectively.}
\label{fig:shutdown}
\end{figure*}

While the dynamics and stage classification presented so far applies to
both crust- and full-configurations, important differences are present
that ultimately determine the final magnetic-field amplification. In
particular, the KHI-driven evolution is very efficient in the
crust-configuration but also ends earlier than for the
full-configuration, thus achieving a smaller magnetic-field
amplification. In addition, the KHI is followed by a longer decay stage
and a shorter turbulent amplification (see red solid line in the left
panel of Fig.~\ref{fig:total_energy}). As a result, when the winding
stage starts for the crust-configuration, the magnetic energy is about
one order of magnitude smaller than in the full-configuration; since the
growth-rates are comparable in the two topologies, this difference
remains unchanged till the end of the simulations. Also worth noting is
that the turbulent amplification is smaller in the crust-configuration as
the result of two combined effects. First, a smaller amount of magnetised
matter is subject to turbulent motions. Second, because the magnetised
matter is only at lower densities (\ie $\rho \lesssim 10^{13}\,{\rm
  gm/cm}^3$) and near the stellar surface, a good portion of it is shed
in the external medium before it can be amplified further. These
  ``winds'', that are in good part (but not exclusively) magnetically
  driven, are the result of large deviations from a magnetohydrostatic
  equilibrium at the remnant surface and are made more violent by the
  presence of comparatively stronger magnetic fields in the crust
  configuration. By contrast, the amplification in the
full-configuration can benefit from larger volumes of magnetised matter
and smaller losses of magnetized matter at the surface, where the
  pressure gradients are comparatively smaller.

The right panel of Fig.~\ref{fig:total_energy} gives us the opportunity
to discuss in more detail the turbulent magnetic-field amplification
during the first two stages of the evolution and thus for
$t-t_{\mathrm{mer}} \lesssim 2\,\mathrm{ms}$. In particular, shown in the
upper parts is the evolution of $E_B$ for the same configurations in the
left panel, while the lower parts report the corresponding normalized
growth-rate $\dot{E}_B/E_B$. First, an exponential growth is present at
the onset of magnetic-field amplification, which then reaches its maximum
value in $\lesssim 0.2 \,\mathrm{ms}~(0.7)$ for the crust- (full-)
configuration. While the maximum growth-rates of the two configurations
are comparable for the same resolution, the changes in $E_B$ are
considerably more rapid in the crust-configuration, as it is natural to
expect since the magnetic energy involved in the KHI amplification is
confined to a smaller volume and nearly all of the initial
electromagnetic energy participates in the amplification process as the
KHI develops. After the maximum of $\dot{E}_B/E_B$ has been reached, the
subsequent evolution is qualitatively very similar in all
configurations. In particular, the growth-rate decays until it reaches
negative values and the KHI-driven stage ends. During this decay, very
short variations of the growth-rate take place with a period of
$\simeq0.25 - 0.35\,{\rm ms}$; considering an eddy-rotation velocity of
$\sim 0.15\,c$, this variation yields eddy lengthscales of $\sim
1.8-2.5\,{\rm km}$, which match well the size of the largest eddies
produced across all simulations. This suggests that the decay in the
growth-rate is related to the largest eddy turnovers in the KHI-driven
stage and the periodic increase/decrease of the growth-rate is the result
of the amplification and subsequent dissipation of large-scale flux tubes
that are dragged into rotation by the largest eddies. Furthermore, due to
periodic bounces of the two cores with a period of $\sim
0.77\,\mathrm{ms}$, the shear layer between the stars is turned over at
half of this period and drives the production of the largest eddies.
This dynamics highlights that a persistent amplification requires the
development of a fully turbulent stellar interior. While this is almost
inevitable in the case of the full-configuration, where magnetized
material is present across the HMNS, continued amplification is difficult
for the crust-configuration, where turbulent amplification is still
present but it is accompanied by the mixing of magnetized and
unmagnetized material. Such mixing, in addition to magnetized-matter
losses at the HMNS surface, reduces the growth-rate and favours
dissipation.

Figure~\ref{fig:shutdown} offers an alternative means of comparing the
full- (top row) and crust-configurations (bottom row) in the KHI-driven
stage by examining representative quantities at $t-t_{\mathrm{mer}} =
0.52~\mathrm{ms}$. From left to right we report: $|B|/[10^{15}\,{\rm
    G}]$, the kinematic vorticity $\left|\rho
\omega^{xy}_{\perp}\right|:= \rho h^{x}_{\phantom{x}\mu}
h^{y}_{\phantom{y}\nu} \omega^{\mu \nu}$, and the amplification source
term, $s_{\mathrm{mag}} := b^{\mu}b^{\nu}\sigma_{\mu \nu} -
b^2\Theta/6$. Here $h^{\mu}_{\phantom{\mu}\nu}$ denotes the projector
tensor onto spatial hypersurfaces in the 3+1
formalism~\citep{Rezzolla_book:2013}, $\omega^{\mu \nu}$ the kinematic
vorticity, $b^{\mu}$ the magnetic field in the fluid-frame, $\sigma_{\mu
  \nu}$ the shear tensor, and $\Theta$ the expansion scalar~\citep
[see][for definitions]{Chabanov2021}. The quantity $s_{\mathrm{mag}}$ is
the general-relativistic counterpart of the quantity employed by~\citep
{Obergaulinger10} to measure the strength of local sources
($s_{\mathrm{mag}} >0$) and sinks ($s_{\mathrm{mag}}<0$) of
electromagnetic energy.

First, when comparing $|B|$ between the two configurations (left column),
it is apparent that the largest values of the magnetic-field strength can
be found in the crust-configuration and that these are reached very close
to the putative surface of the HMNS (marked with a green contour line at
$10^{13}\,{\rm gm/cm}^3$), while the inner parts are essentially devoid
of magnetic field with the exception of those regions that belonged to
the stellar surface. Second, when looking at the vorticity (middle
column), it also clear that the difference between the two configurations
is very small and that in both cases the turbulent motion is concentrated
in the low-density regions and falls-off rapidly when moving towards the
stellar interior where, again, the turbulence is present only in the
regions which were on the stellar surface. Finally, when comparing the
sources and sinks of magnetic energy (right column), it is possible to
realize that in both cases the sources are larger than the sinks (hence
the amplification), but also that a large portion of the sources in the
crust-configuration are just about to be lost at the HMNS surface. It is
exactly the shedding of this precious, highly magnetized material that
will quench the further amplification of the crust-configuration and
ultimately lead to smaller magnetic fields. The emission of
  neutrinos may further increase the strength of these winds, thus
  additionally reducing the potential of magnetic-field amplification in
  crust configurations.

\begin{figure}
\includegraphics[width=0.99\columnwidth]{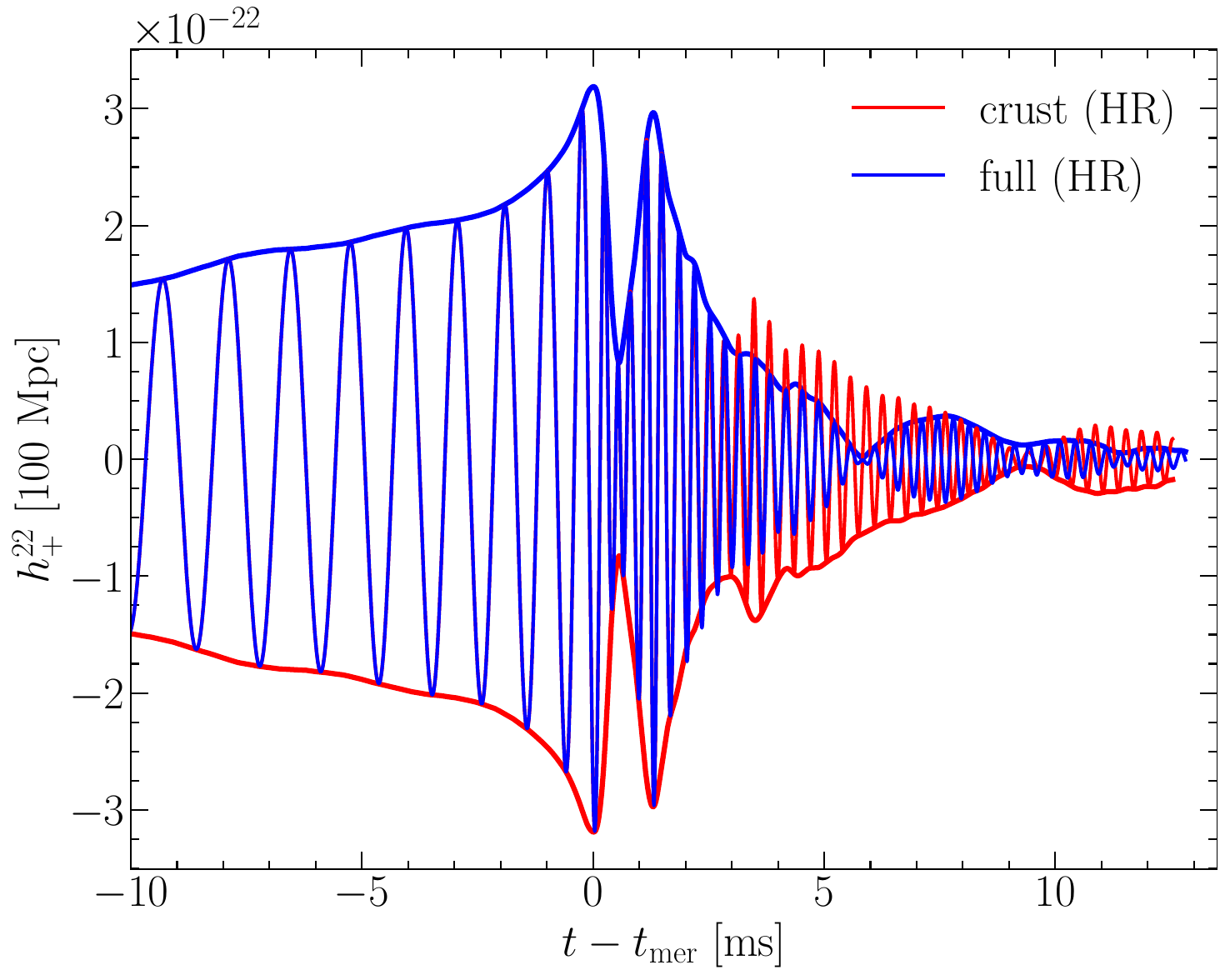}
\caption{Gravitational-wave strain in the $\ell=2,m=2$ mode of the
  $+~$--polarization extracted at $\sim 740~\mathrm{km}$ for a source at
  and normalized to a distance of $100\,{\rm Mpc}$ and for the two
  configurations. Thick solid lines report the corresponding amplitudes.}
\label{fig:gws}
\end{figure}

We conclude our analysis by discussing the impact that the different
magnetic-field topologies have on the emitted gravitational-wave signal,
which we report in Fig~\ref{fig:gws}. Note that the signals from the two
configurations are indistinguishable during the inspiral and very similar
over the first two milliseconds after the merger, when the KHI is most
active. However, after $t-t_{\mathrm{mer}} \gtrsim 2\,\mathrm{ms}$ the
waveforms differ considerably both in phase and amplitude, with the
latter being smaller for the full-configuration. This is because the
production of magnetic fields comes at the cost of the kinetic energy of
the system in the full-configuration. In addition, the strong
amplification of the magnetic field in the full-configuration reduces the
$m=2$ deformations leading to a more axisymmetric HMNS in the
full-configuration and hence to a weaker gravitational-wave signal.

\section{Conclusions}
\label{sec:conclusions}

Motivated by a commonly assumed scenario in which magnetic fields in
neutron stars are confined to a small region near the stellar surface, we
have performed high-resolution, global, and high-order GRMHD simulations
of merging BNSs with different initial magnetic-field topologies. In
particular, while keeping the magnetic energy the same, we have
investigated the impact of a purely \textit{crustal} magnetic field and
contrasted it with the standard configuration consisting of a dipolar
magnetic field filling the whole star.

The use of high spatial resolution, high-order methods, and realistic
initial magnetic fields has allowed us to highlight the presence of four
distinct stages in the evolution of the magnetic field.  While these
stages are common to both configurations, and although the
crust-configurations are more efficient in generating strong magnetic
fields during the KHI-driven stage, we find that crust-configurations
fail to achieve the same level of magnetic-field amplification as their
full counterparts. We attribute this behaviour to the lack of magnetized
material in the neutron-star interiors that can be used for further
amplification and to the losses at the stellar surface of highly
magnetized matter that afflicts the crust-configurations. As a result,
by the end of our simulations the magnetic energies in the two
configurations differ by a bit more than an order of magnitude and the
gravitational-wave signal in the full-configuration is $\sim 50\%$ weaker
than in the crustal counterpart as a result of a larger degree of
axisymmetry.

Inevitably for global fully general-relativistic simulations of this
type, ours also suffers from resolutions that, while very high and
computationally expensive, are insufficient to capture a fully convergent
behaviour during the KHI-driven exponential growth. However, by
performing simulations with different resolutions we have verified that
the behaviour presented here is only quantitatively modified by
resolution and that the qualitative features remain unaltered. Hence, we
expect our results to provide a reasonably accurate description of the
magnetic-field amplification for stars with realistic initial magnetic
fields.

Our findings have at least two important implications. First, future
observations providing evidence for the presence of magnetar-strength
magnetic fields in the merger remnant will represent a clear indication
that the magnetic-field topology before merger could not have been a
purely crustal one. Second, since the main difference between the two
configurations considered here is represented by the volume fraction
endowed with magnetic field, it is possible to correlate the post-merger
dynamics -- both in terms of gravitational-wave emission and in ejected
matter -- to the fraction of stellar volume that is magnetized. We leave
these investigations to future work.

\begin{acknowledgments}

  We thank Vasilis Mpisketzis for useful discussions. Partial funding
  comes from the GSI Helmholtzzentrum f\"ur Schwerionenforschung,
  Darmstadt as part of the strategic R\&D collaboration with Goethe
  University Frankfurt, from the State of Hesse within the Research
  Cluster ELEMENTS (Project ID 500/10.006), by the ERC Advanced Grant
  ``JETSET: Launching, propagation and emission of relativistic jets from
  binary mergers and across mass scales'' (Grant No. 884631) and the
  Deutsche Forschungsgemeinschaft (DFG, German Research Foundation)
  through the CRC-TR 211 ``Strong-interaction matter under extreme
  conditions''-- project number 315477589 -- TRR 211. LR acknowledges the
  Walter Greiner Gesellschaft zur F\"orderung der physikalischen
  Grundlagenforschung e.V. through the Carl W. Fueck Laureatus Chair. The
  simulations were performed on HPE Apollo HAWK at the High Performance
  Computing Center Stuttgart (HLRS) under the grant BNSMIC.  ERM
  gratefully acknowledges support as the John A. Wheeler Fellow at the
  Princeton Center for Theoretical Science, the Princeton Gravity
  Initiative and the Institute for Advanced Study.  ERM acknowledges
  support through the Extreme Science and Engineering Discovery
  Environment (XSEDE) through Expanse at SDSC and Bridges-2 at PSC
  through allocations PHY210053 and PHY210074. ERM further acknowledge
  supported by Princeton Research Computing, a consortium of groups
  including the Princeton Institute for Computational Science and
  Engineering (PICSciE) and the Office of Information Technology's High
  Performance Computing Center and Visualization Laboratory at Princeton
  University.
\end{acknowledgments}

%


\software{\texttt{Einstein Toolkit} \citep{Loffler:2011ay},
          \texttt{Carpet} \citep{Schnetter:2003rb},
          \texttt{FIL} \citep{Most2019b},
          \texttt{FUKA} \citep{Papenfort2021b},
          \texttt{Kadath} \citep{Grandclement09},
          \texttt{CompOSE} (\url{https://compose.obspm.fr})
}



\appendix

\section*{Simulations Details and Impact of resolution}

\begin{figure*}
\includegraphics[width=0.99\textwidth]{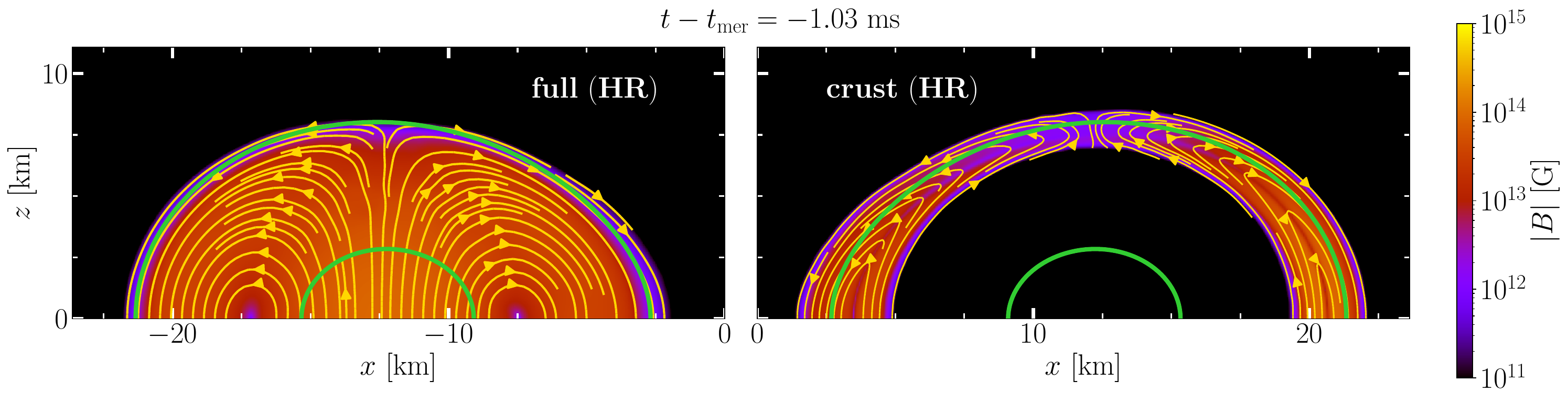}
\caption{Distributions on the $(x,z)$ plane of the magnetic-field
    strength $|B|$ from the HR simulations soon before
    merger. Magnetic-field lines are shown in yellow, while the left and
    right panels refer to the full- and crust-configuration,
    respectively. }
\label{fig:initial}
\end{figure*}

In what follows we provide additional details on the features of the
simulations reported in the main text, as well as a discussion of how the
amplification differs when considering lower resolutions.  In particular,
we recall that the simulations have been performed with the \texttt{FIL}
code, which employs fourth-order accurate finite-difference stencils in
Cartesian coordinates for the evolution of the constraint damping
formulation of the Z4 formulation of the Einstein
equations~\citep{Bernuzzi:2009ex, Alic:2011a}, while the equations of
GRMHD are solved with a fourth-order high-resolution shock-capturing
scheme~\citep{DelZanna2007}, together with a vector-potential-based
constrained transport scheme inherited from the open-source code
\texttt{IllinoisGRMHD} \citep{Etienne2015}. On the other hand, initial
data is computed using the \texttt{FUKA} codes~\citep{Papenfort2021b}
where the equal-mass binaries are chosen to be irrotational with a total
ADM mass of $\sim 2.55~M_{\odot}$ initialized at a separation of $\sim
30~M_{\odot}\approx 44\,\mathrm{km}$. The computational grid has outer
boundaries at $1000\,M_{\odot}\simeq 1476\,{\rm km}$ in the three spatial
directions and we employ a $z$-symmetry in the equatorial plane. We use
seven refinement levels with factor of two refinement; the last level,
which has a width of $32~M_{\odot}$, is added prior to merger, when the
separation between the ``barycenters'' of the two stars is
$\lesssim9~M_{\odot}$.

Table~\ref{tab:models} reports the smallest employed cell size $\Delta
x$, the parameters of the initial magnetic field, the initial maximum
magnetic-field strength $|B|_{\mathrm{max}}^{\mathrm{seed}}$. Also
reported are the times characterizing the different stages of our
simulations, namely, the end of the KHI-driven stage
$t_{\mathrm{end}}^{\scriptscriptstyle \mathrm{KHI}}$, the end of the
decay stage $t_{\mathrm{end}}^{\scriptscriptstyle \mathrm{D}}$, and the
end of the turbulent-amplification stage
$t_{\mathrm{end}}^{\scriptscriptstyle \mathrm{TA}}$.

\begin{deluxetable}{lrrrrrrrrrrr}
\tablenum{1} \tablecaption{Characterising information on the different
  models explored in this work. We show the cell size on the highest
  refinement level, the seed magnetic-field parameters employed to
  initialize the magnetic field, the maximum magnetic-field strength
  after initialization, $|B|_{\mathrm{max}}^{\mathrm{seed}}$, as well as
  the end time of the different stages discussed in the text, i.e.
  $t_{\mathrm{end}}^{\scriptscriptstyle \mathrm{KHI}}$,
  $t_{\mathrm{end}}^{\scriptscriptstyle \mathrm{D}}$ and
  $t_{\mathrm{end}}^{\scriptscriptstyle \mathrm{TA}}$,
  respectively. }
\label{tab:models}
\tablehead{
Model & $\Delta x$ & $R_{\rm in}$ &
$A_b$ & $g_w$ & $g_r$ &
$p_{{\rm co}}$ & $n$ &
$|B|_{\mathrm{max}}^{\mathrm{seed}}$ &
$t_{\mathrm{end}}^{\scriptscriptstyle \mathrm{KHI}}-t_{\mathrm{mer}}$ & 
$t_{\mathrm{end}}^{\scriptscriptstyle
\mathrm{D}}-t_{\mathrm{mer}}$ &
$t_{\mathrm{end}}^{\scriptscriptstyle \mathrm{TA}}-t_{\mathrm{mer}}$ \\
& $[\mathrm{m}]$ & $[M_{\odot}]$ & &
$[M_{\odot}^{-2}]$ & $[M_{\odot}]$ &
$[M_{\odot}^{-2}]$ & & $[\mathrm{G}]$ &
$[\mathrm{ms}]$ & $[\mathrm{ms}]$ &
$[\mathrm{ms}]$ }
\startdata
full (HR)  & $70$  & $0$ & $0.028$ & $0$ & $0.0$ & $1.0\times10^{-8}$ &
$1.00$ & $1.04\times10^{14}$   & $2.46$ & $3.48$ & $7.13$ \\
full (LR)  & $105$ & $0$ & $0.028$ & $0$ & $0.0$ & $1.0\times10^{-8}$ &
$1.00$ & $1.04\times10^{14}$   & $1.77$ & $5.15$ & $10.1$ \\
crust (HR) & $70$  & $5$ & $0.131$ & $4$ & $6.1$ & $1.0\times10^{-8}$ &
$0.85$ & $2.36\times10^{14}$ & $0.90$  & $3.59$ & $5.09$ \\
crust (LR) & $105$ & $5$ & $0.120$  & $4$ & $6.1$ & $1.6\times10^{-7}$ &
$0.85$ & $2.26\times10^{14}$ & $1.82$ & $3.54$ & $11.5$ 
\enddata
\end{deluxetable}

\begin{figure*}[ht!]
\includegraphics[width=0.99\textwidth]{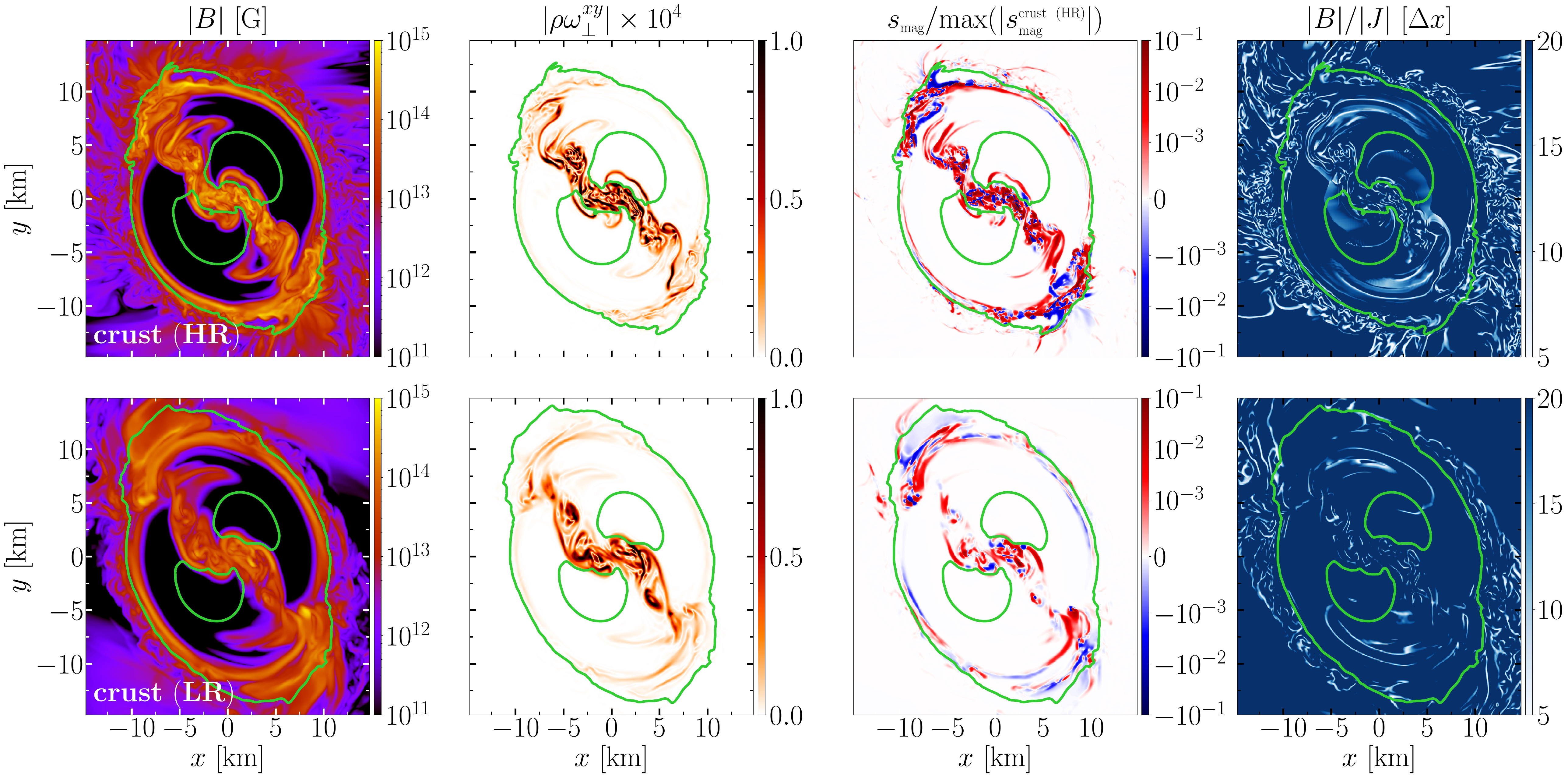}
\caption{The first three columns from left as the same as in
  Fig.~\ref{fig:shutdown}, while the fourth column reports the
  characteristic length-scale of the magnetic field, \ie $|B|/|J|$, when
  expressed in units of resolution spacing $\Delta x$.}
\label{fig:shutdown_appendix}
\end{figure*}

In analogy with, Fig.~\ref{fig:normB}, we use Fig.~\ref{fig:initial}
  to offer a more intuitive view of the different topologies of the
  magnetic field also on the $(x,z)$ plane. In particular, the figure
  reportes the distributions of the magnetic-field strength $|B|$ from
  the HR simulations soon before merger. Magnetic-field lines are shown
  in yellow, while the left and right panels refer to the full- and
  crust-configuration, respectively; note that in both cases the magnetic
  field is purely poloidal.

Next, and in addition to what is already presented in
Fig.~\ref{fig:total_energy}, we discuss how the resolution impacts our
results by contrasting the LR and HR version of the crust-configuration
(we do not concentrate here on the full-configurations as these show very
similar behaviour for both resolutions so that the discussion on full
(HR) is applicable to full (LR), also these have been discussed in a
number of papers, \eg see~\citet{Most2019b} for a case with lower
resolutions and higher magnetic-field strengths compared to the
simulations in this work). Bearing in mind that the resolutions employed
here do not allow for a rigorous convergence study, the overall evolution
of the LR and HR-simulations provide evidence of being numerically
consistent, \ie that the errors should decrease with resolution. More
specifically, Fig.~\ref{fig:shutdown_appendix} compares the HR (top row)
at $t-t_{\mathrm{mer}}=0.52~\mathrm{ms}$ with the LR-simulations (bottom
row) at $t-t_{\mathrm{mer}}=0.59~\mathrm{ms}$, where the different times
are due to the difference in phase evolution due to the different
resolution.  The first three columns from the left report the same
quantities as shown in Fig. \ref{fig:shutdown}. The fourth column, on the
other hand, reports the characteristic length-scale of the magnetic
field, \ie $|B|/|J|$, when expressed in units of resolution spacing
$\Delta x$ \citep[see Eq.~(15) of][]{Obergaulinger10}. Here,
$J:=\sqrt{J^{i}J_{i}}$ and the spatial current is estimated as
\begin{align}
  \label{eq:current}
J^{i} \simeq \frac{1}{4\pi \alpha}\epsilon^{ijk}D_{j}\left(\alpha B_k\right)\,.
\end{align}
Expression~\eqref{eq:current} follows from the fact that in ideal-GRMHD
the spatial components of the electric field can be expressed as $\alpha
E_i = -\epsilon_{ijk} \left(v^{j}+\beta^{j}\right)B^{k}$, so that for
nonrelativistic flows, $E_i \sim \mathcal{O}(v_i) \ll 1$ and the current
reduces to Eq.~\eqref{eq:current}.

Starting from the first column of Fig.~\ref{fig:shutdown_appendix} it is
possible to realize that the HR simulation loses more magnetized material
across the HMNS surface than the LR counterpart. This loss is most
significant at both ends of the turbulent shear layer where also
dynamical ejecta will originate from. The second column compares the
densitized vorticity and shows that, as expected, more vorticity is
present in the HR simulation and that the differences are more marked at
both ends of the turbulent interface. The third column shows that stronger
sinks (higher in magnitude but negative) are present at the ends of the
turbulent interface in the HR simulation. Finally, the fourth column
clearly shows that the HMNS in the HR simulation is surrounded by a
``cloud'' of low-density and magnetized material with a very small
characteristic length-scale, which is not visible in the LR
simulation. All in all, Fig. \ref{fig:shutdown_appendix} illustrates how
higher resolution for the crust-configuration can lead to enhanced
shedding of highly magnetized material at the HMNS surface and hence that
the magnetic-field amplification is bound to be smaller as opposed to the
full-configuration.

\bibliographystyle{aasjournal}
\bibliography{aeireferences}

\end{document}